\documentclass[twoside]{article}
\usepackage{amsmath,amssymb,times}
\usepackage[latin1]{inputenc}
\usepackage[english,francais]{babel}
\usepackage{graphicx}
\begin{document}

\date{}
\title{From Galilean Relativity to Quantum Gravity: Three Issues, $c$, $\hbar$, $G$, concerning Time, Matter and Inertia}
\author{Jean-Pierre Provost}
\maketitle \emph{To be published in the proceedings of the $11^{th}$ \textit{Fundamental Frontiers of Physics} symposium, Paris from the 6th to the 9th of Jully, 2010}

\begin{abstract}
An odd look at "standard" physics (Galileo, Newton, Einstein, Dirac) leading to a radical change of our concept of inertial motion and to new heuristic approaches of gravitation and the cosmological constant".
\end{abstract}

\section{Introduction: $c$, $\hbar$, $G$, philosophy and outline}
The philosophy underlying this presentation (which is a brief summary of \cite{JPP2011a}) is that physics, in essence, is relativistic, quantum, and (may be) gravitational. This does not mean only that  $c$, $\hbar$, and (may be) $G$,are fundamental constants, but also that they play a major role in our way of thinking. For example, \textit{Einstein's} 1907 relation $E_0=mc^2$ does not only precise the energetic effect $(\Delta m)c^2$ of a mass defect ; it also tells us that Newton's mass and rest energy are the same thing, although these notions have been historically introduced in quite different contexts. In the same way, the true lesson of \textit{De Broglie's} 1923 relation $mc^2=h\nu_0$ is that mass is nothing else than a frequency ; the more familiar relation $h\nu=\Delta E_0$ describing atomic or nuclear radiative transitions is a consequence of it. Finally, \textit{Planck's} 1900 observation that $Ghc^{-3}$ is a surface is presently often considered as an indication that quantum gravity will seriously modify our conception of space and time. 
\par As a consequence of this philosophy, one must find fundamental and simple issues, starting from Galileo-Newton's physics and leading to the idea that it is quite natural to set $c=1$, $\hbar=1$, and (may be) $G=1$. Obviously, these issues will deal with the basic and strongly connected notions of time, matter and inertia, which special relativity (S.R), quantum physics (Q.P) and general relativity (G.R) have deeply revised in the last century. We formulate them through three apparently odd and naive questions (examined in sections 2,3,4): (1) \textit{How is it possible, starting from Galileo's discorsi to define time ... and recover Einstein's 1905 relation} $\tau =t(1-v^2/c^2)^{1/2}$ ? (2) \textit{How is it possible to introduce mass "\`a la Maupertuis" ... and think of it as a frequency} ($mc^2=h\nu_0$)? (3) \textit{How can this imply Einstein's 1911 formula} $\tau \simeq t(1+\phi/c^2)$ \textit{and the identity of inertial and gravitational masses}? (1) is since long our personal way of understanding the logical necessity of S.R \cite{JPP1980}. (2) originated from an old epistemological reflexion on the physical meaning of Action : since inertia is our reference in physics (cf. inertial frames), and since fundamental laws in physics are derived from Least Action Principles (L.A.P), Action has to do with non inertia. It recently led to the idea that we must radically change our Newtonian conception of "what is inertial and what is not" \cite{JPP2007}. In short, we claim that inertia corresponds to motion at velocity $c$ (realized by zero mass particles) and that, in contradistinction, rest or motion at velocity $v<c$ (realized by massive ones) imply non inertial processes with frequency "m". Finally (3) is a first heuristic application of this new point of view to gravitation. 
\par Of course, this philosophy must be mathematically supported. We emphasize in section 5 that the spinor formalism, with the notion of conjugation, is well adapted. Indeed, \textit{mass can be introduced in Weyl equations as a transition frequency between conjugate spinors and G.R with a cosmological constant may follow from the definition of a spinor connection involving them}. In conclusion (section 6) we recall that physics is impossible without references and we briefly sum up the old, present, and (what we think to be) future ones. 

\section{From Galileo's discorsi to Special Relativity : "defining" time}
In a well known poetic text, Galileo has defined boats in inertial motion, i.e. boats where the flight of flies and physical laws are euclidean invariant ; he also noted that this invariance extends to boosts. But although he was the first to use clocks (pulse, pendulum) in physics, he did not care to discuss time. Following his metaphor, let us consider two sailors first at rest with respect to the boat (frame $R$) at same position $A$, then climbing its mast, and remaining at rest at its top $B$ (figure 1). Seen from any other inertial frame $R'$, their motions which both concretize in $R$ the geometrical translation $\vec{R}=\vec{AB}$ can be called asymptotically inertial motions since, asymptotically, the velocity is unchanged. These motions, which are the simplest ones after inertial motions, allow to introduce time in a relativistic (i.e. frame idependent) way. Indeed, let $A'$ and $B'$ be the extremities in $R'$ of their inertial parts, and let $\vec{R'}=\vec{A'B'}$ correspond to the non inertial one. Then asserting that the sailors have climbed the mast in the same time is clearly equivalent to the geometrical relation  "$\vec{R'_1}=\vec{R'_2}$ in any frame". 
\par In order to express this idea in mathematical terms, let us consider two frames related by an infinitesimal boost $\vec{\epsilon}$, $\vec{\epsilon}$ being just a group parameter with vectorial character ; one expects $\vec{R'}=\vec{R}+\vec{\epsilon} \times$ (scalar quantity), and because time has been introduced by "$\vec{R'_1}=\vec{R'_2}$ whatever $\vec{\epsilon}$", the scalar quantity defines the time $T$ necessary for the sailors to climb the mast in $R$. In $R'$, one also expects $T'=T+\vec{\epsilon}.$(vectorial quantity) and "$T'_1=T'_2$ whatever $\vec{\epsilon}$" since our introduction of time is intended to be relativistic. The simplest hypothesis is that the vectorial quantity is proportional to $\vec{R}$ (remind that already $\vec{R_1}=\vec{R_2}$). Finally we get the transformations $\vec{R'}=\vec{R}+\vec{\epsilon} T$, $T'=T+\mu\vec{\epsilon}.\vec{R}$ which imply $T^2-\mu\vec{R}^2$ invariant. Causality ($\mu\ge 0$), genericity  ($\mu\neq 0$) and simplicity $\mu= 1$ lead to S.R with $c=1$, and Michelson's experiment tells us that this invariant velocity is that of light. 
\par The simplest clock with respect to this introduction of time is clearly a system of two mirrors with light going to and fro between them ; it could not be a pendulum! If $T_0$ is the period in the mirrors frame, the ACTIVITY of the clock, defined by the number of tic-tac (bounces of light on the mirrors) is 
$$
A=\tau / T_0    \hspace{2cm} \rm{with}  \hspace{2cm}  \tau = \int dt (1-v^2)^{1/2}.
$$

Figure 2 illustrates it for $v=0$ and $v$ constant. If the clock moves from $A$ to $B$ in a given time in some frame, its activity is maximum when $\vec{v}$ is constant (cf. Langevin's twins) and it goes to zero when $v(t)$ comes close to $c=1$  ( the "tomb of time"). Let us remark finally that, in practice (cf. atomic clocks), such a clock needs matter in order to control the frequency of the stationary wave inside the mirror cavity ($h\nu=\Delta E_0 = \Delta m c^2$). This leads to the question : is not matter itself a clock? 

\section{From Maupertuis' action to Quantum Physics : "defining" mass and inertia.}
Wanting to unify Optics and Mechanics, Maupertuis proposed in 1744 to replace Fermat's principle "$\Sigma l / v$ min" by "$\Sigma v l $ min" (which leads to Descartes' law $v_1 sin i_1 = v_2 sin i_2$ instead of $ sin i_1 / v_1 =  sin i_2 / v_2$). In 1746, he extended it to "$\Sigma m v l $ min" in order to account for the conservation of momentum in collisions. Following him, we consider the elastic collision of two bodies $i=1,2$, with the initial $A_i$ and final  $B_i$ positions and the corresponding time coordinates being fixed (figure 3). In absence of collision, the bodies would have travelled from  $A_i$ to $B_i$ at constant speed, obeying "$\tau_i$ max". If one imposes them to collide, one cannot satisfy both "$\tau_1$ max" and "$\tau_2$ max", but one knows from experience that the motion of the largest mass is the closest to an inertial motion. So, it is natural to "ponderate" the conditions for $\tau_1$ and $\tau_2$ and to introduce mass by postulating "$m_1 \tau_1 + m_2 \tau_2$ max" (a condition which additionally gives the position and time of collision). This compromise which agrees with Newton's still present conception of inertial mass (resistance to a change of velocity) leads very simply to the conservation law of the relativistic energy momentum (because $md(T^2-\mu\vec{R}^2)^{1/2}=EdT-\vec{p}.d\vec{R}$). 
\par Remarkably, the L.A.P  "$\Sigma m_i \tau_i$ max" also works when the particles $i$ appear or disappear, as in $n \rightarrow p+e^-+\bar{\nu}$, i.e. very far from the case of elastic collisions at the origin of our intuitive idea of mass. If each quantity $ m_i\tau_i$ has still to do with inertia and non inertia, which is our credo, it can no longer concern the global motion of the mass $m_i$. We claim that it concerns the internal dynamics of $m_i$, in the same way as Einstein has told us that mass, being rest energy, is determined by this dynamics. More precisely, we make the following ANSATZ : " {\it The quantity} $A=m \tau$ {\it is a number which counts the "internal" non inertial events associated with the object of mass m}" ; as a COROLLARY : "{\it Inertia corresponds now to $A=0$, i.e. to motions at velocity $c=1$ realized by $m=0$ particles, and the simplest non inertial processes (events) are the annihilation/creation (A/C) of such motions}". Clearly, this ansatz is motivated by the analogy between $A=m \tau$ and the ACTIVITY $A=\tau / T_0$ of the clock of section 2, and by figure 2. But one must not confuse it with a model of particles. It is just a quantum, or at least semi-classical, PARADIGM of mass. In Q.P language, $A=mc^2\tau/ \hbar$ is the quantum phase and the above conservation law is a phase matching. 
\par It is easy to put this ansatz on a more formal ground. Let in $d=1$ dimension $(\partial_t+\partial_z)f=0$ and $(\partial_t-\partial_z)b=0$ ($f=$ forwards, $b=$backwards) be the equations associated with inertia. Then, their negative and positive frequency solutions $f_\mp \propto \exp \pm iE(t-z)$ and $b_\mp$ are the amplitudes corresponding to the A/C of motions at $c=1$ and $c=-1$. Mass as introduced above describes the coupling between $b_-$ and $f_+$ and between $f_-$ and $b_+$ since it is a frequency of velocity change from $+c$ to $-c$. This leads us to introduce it through the equations $(\partial_t+\partial_z)f=mb^*e^{i \alpha}$, $(\partial_t-\partial_z)b=mf^*e^{i \beta}$. Their dispersion relation (for $f$, $b^*$, proportional to $ \exp - i(Et-pz)$) being $E^2=p^2-m^2 \exp  i(\alpha-\beta)$, one needs  $\alpha=\beta+\pi$ for $E$ and $p$ to be real ; then one gets $v=p/E=(|f|^2-|b|^2)/(|b|^2+|f|^2)$, i.e. the velocity of the mass $m$ (group velocity of the plane wave) is also the mean velocity obtained from the value +1 and -1 and probabilities proportional to $|f|^2$ and $|b|^2$ ; a simple calculation shows that  $|f|^2$ and $|b|^2$ are proportional to the times of duration of the motions c=+1 and c=-1 on figure 2. In the non relativistic limit ($v<1$ or $|f| \simeq |b|$), one recovers of course the Schr\"{o}dinger equation for a free particle. The generalisation to $d=3$ is sketched in section 5 ; then $f$ and $b$ appear to be the components of a spinor, which shows that this approach is quite different from De Broglie's one in his thesis (De Broglie added plane waves propagating at $c=+1$ and $c=-1$, say $f_-$ and $b_-$). 
\par Finally, let us make three general comments in relation to this quantum introduction of mass. The first one is that time itself has become "quantum". Indeed, since $\tau = A$ if $m=1$, {\it time is the activity of some arbitrary chosen mass} ; as a consequence, matter is responsible for its flow, a point of view which Leibniz, but not Newton could have shared. The second one concerns the {\it experimental access to non inertia} ; in the same way as Compton's diffusion on a massive particle directly gives its characteristic time of non inertia $\tau_c=m^{-1}$ through $\lambda'-\lambda=m^{-1}(1-\cos \theta)$, deep inelastic scattering introduces the characteristic times $(mx)^{-1}$ $(x<1)$ through $\lambda'-\lambda=(mx)^{-1}(1-\cos \theta)$; then the theory (Feynman's model of zero mass partons for hadrons) gives the contributions of the different partons (quarks, antiquarks, gluons ...) to the processes occuring with the frequency $mx$ ("Compton analysis" of particles also called infinite momentum frame analysis). The third comment is that the zigzag motion of figure 2 suggests that non inertia can also be associated with a swept area in space time. Its order of magnitude being $s=(m \tau) \times m^{-2}$ (remind that $T_0=m^{-1}$), it gives a physical interpretation to the affine parameter defined in S.R (and in G.R) by $p^\mu =dx^\mu/ds$ and connecting dynamics and kinematics. This remark on {\it non inertia and swept area} allows to characterize geodesics (even in G.R) not only by "$\tau$ or $A$ max", but equally by "s max" \cite{JPP2011b} (figure 4). It will offer the opportunity to introduce $G$ and gravitation in the next section. 

\section{Revisiting Newtonian gravity : $G$ as a surface.}
Since mass is a frequency of non inertia, there is a simple way to understand why the masses of isolated systems at rest add (still a naive question): they add because activities (number of non inertial events) do, in the same way as the frequencies of independent Poisson processes. But gravitation tells us that it is not quite true ; so doing, we count too many events (attractive character of gravitation). Indeed, for two bodies at rest, distant from $r$, one has in the Newtonian limit (and mass being rest energy):
$$
M \Delta t = m_1 \Delta t +m_2 \Delta t -(m_1 \Delta t)m_2G/r \hspace{0,5cm} (c=1).
$$
It seems to indicate that \textit{non inertial events at points separated by $r$ cannot be distinguished if they occur during a time interval} $\delta t =G/r$. It is important to note that, whereas the relation between $m_1$, $m_2$ and $M$ is "classical", thinking of $G/r$ as a time interval is quantum ($\delta t =G \hbar/rc^4$ if $c \neq 1$, $\hbar \neq 1$). This uncertainty is not a surprise if one calls for Q.P and G.R : let a clock $m$ stay at origin and send to $r$ signals as short as possible for synchronisation ; then from $(\Delta m c^2)\Delta t \simeq \hbar$ (Heisenberg), $\Delta \varphi =G \Delta m/r$ (Newton) and $\tau=t(1+\varphi/c^2)$ (Einstein), one gets $\delta \tau =\Delta t \Delta \varphi /c^2=\delta t$. But our goal is different ; we want to justify $\delta t$, at least heuristically, and deduce Einstein's proper time; Newtonian gravitation then follows.
\par In order to justify $\delta t$, we recall that a mass at rest is the center of non inertial events. Rest must certainly also imply some non inertia, even in the absence of a test mass. This leads us to suppose that if a signal is sent from origin, its detection at $r$ is necessarily accompagnied by a coming backwards of it (figure 5). If the associated swept area is $G$ (i.e. $G \hbar c^{-3}$) up to a constant, then the constant can be chosen such that the detection needs the above time $\delta t$. If a mass at origin, i.e. a clock helps as a reference to define rest at point $r$, it requires a fraction $(mc^2\hbar^{-1})\delta t$ of the universal time $t$. The remaining time at disposal for physics (proper time) at this point is $\tau = t(1-mc^2\delta t/\hbar)=t(1-Gm/rc^2)$ ; it extends to Einstein's formula $\tau = t(1+\varphi/c^2)$ with $\Delta \varphi = 4 \pi G \varrho$ if rest is defined with respect to a Newtonian distribution of masses. The remarkable fact concerning this somewhat Machian reasoning is that, although $\hbar$ disappears in the expression of $\tau$, gravitation is a quantum effect because of the role played by $G\hbar c^{-3}$ as a surface. In addition, since gravitation appears to be (in the Newtonian limit) a correction to the naive addition of activities (non inertiae) for a set of bodies, the equality of inertial and gravitational masses (the oldest law in physics) becomes a triviality. \par Finally, let us note that \textit{the above quantum  link between non inertia and surface seems to be a general feature of gravitation} \cite{JPP2009}. A first example is the relation between the typical activity $A \simeq mr$ $(r \simeq Gm)$ of a Schwarzschild black hole and the surface of its horizon $s \simeq r^2 \simeq GA$. (As a side remark $A$ is also the Bekenstein Hawking entropy $S$, and this is not a surprise because, if one imagines that the activity is due to partons emitted and absorbed by the singularity, then $\tau_c=m^{-1} \simeq r/n$ where $n$ is their number, and therefore $S \simeq A \simeq n$). 
A second one, a little bit mysterious, concerns cosmology ; if the past activity of the universe $A=Ma$ is at the origin of its present "surface", i.e. if $a^2=GA$, then this " \textit{fractal crumpled universe}" has a density $\varrho \simeq Ma^{-3} \simeq (Ga^2)^{-1}$ ; (the right order of magnitude in comparison with $G^{-2}$ or $a^{-4}$!). Finally, the Raychaudhuri relation $ds=-4 \pi G (T_{\mu \nu} dx^\mu dx^\nu s)$, which in G.R describes the shrinking of the section of a parallel pencil of light, is of the same type because, for dust matter at rest, the parenthesis is the activity of matter inside the volume swept by the pencil. Of course, these examples or oddities need theoretical support.  

\section{Inertia, non inertia and spinors.}
In 1843, Hamilton introduced his quaternions which elegantly algebrize our 3d geometry in the same way as complex numbers do for plane geometry. For him this introduction was part of a larger program: "It appeared to me ... to regard algebra ... not primarily as a science of quantity ... but rather as a science of order in progression ..., as science of pure time." Whatever the true intentions of Hamilton, we want here to understand his above last words as "science of processes..., science of inertia/non inertia". Today with the development of Q.P, physicists prefer to keep the complex algebra as the fundamental one (description of probability amplitudes, link between A/C processes and negative/positive frequencies ...). The algebra of quaternions is replaced by that of Pauli matrices and the quaternionic formalism by a spinorial one. Spinors (right and left) are the basic objects of physics in the sense that they correspond to the two fundamental conjugate representations of the Lorentz group. In (quantum) particle physics, they allow to write invariant field equations which account for the A/C of particles/antiparticles of helicity $\pm 1/2$ (Weyl) or of spin $1/2$ (Dirac). In (classical) space time geometry, they are used to describe the light cone, and conjugate spinors correspond to opposite spatial directions. \textit{This double role of spinors, together with their dimensionalities $L^{-3/2}$ and $L^{1/2}$ differing by a surface, make them of interest for implementing the above  ideas of inertia $(v=c)$ and non inertia, and for discussing gravitation} (figures 2, 5). 
\par Let us first come back to the introduction of mass in section 3 through the equations  $(\partial_t+\partial_z)f=mb^*$,$(\partial_t-\partial_z)b=-mf^*$; we recall that their solutions for $m=0$ describe the basic non inertial processes. In $d=3$, their generalisation are the Weyl-Majorana equations (W.M. eqs) $(\partial_t+\vec{\sigma}.\vec{\nabla})\psi = m \epsilon \psi^*$ or $(\partial_t-\vec{\sigma}.\vec{\nabla})\psi = m \epsilon \psi^*$ which are related to each other by changing the spinor $\psi$ into its conjugate $\epsilon \psi^*$ $(\epsilon=i \sigma_2)$. For $m=0$ (Weyl eqs.) the solutions still describe A/C of motions at velocity $c=1$ (in a direction $\hat{n}$), but they get a chiral property, in the same way as, in particle physics they describe the annihilation of a neutrino and the creation of an antineutrino, or vice versa. Therefore not only inertia is no longer associated with the classical notion of inertial frames (space moving at $v<c$ without acceleration or rotation) but it has also become chiral ; if this is true the chirality of particles is a consequence of that of inertia (a similar remark may apply to the origin of Fermi statistics).  W.M. eqs. are non standard wave equations (they mix $\psi$ and $\psi^*$) and may prefigure amplitude equations of a future non linear theory. In particle physics one presently prefers to them the Dirac eq. $(\partial_t+\vec{\sigma}.\vec{\nabla})\psi_R= m \psi_L$,  $(\partial_t-\vec{\sigma}.\vec{\nabla})\psi_L= m  \psi_R$, without paying attention to the fact that it reduces to two independent W.M. eqs. (for $\psi_R \pm \epsilon \psi_L^*$). Certainly the notion of (anti)particle has been up to now a fundamental one and it will remain so as long as the distinction between particles and interactions makes sense. But whatabout it at Planck's length? 
\par As well known, spinors (written in some inertial frame) generate the (forward) light cone by $\psi \psi^+ = t(1+\vec{\sigma}.\hat{n})$ and the conjugation $\psi \rightarrow \epsilon \psi^*$ changes $\hat{n}$ in $-\hat{n}$. We claim that more fundamentally \textit{one can define a physical local space time (in absence locally of matter) by the connection}
$$
\delta_X \psi = a^{-1}q\epsilon \psi^* \hspace{1cm} \rm{with} \hspace{1cm} q=dt+\vec{\sigma}.d\vec{r}=e^I_\mu \sigma_I dx^\mu
$$
($e^I_\mu$ is the tetrad field connecting free fall and arbitrary coordinates). Because it couples $\psi$ and $\psi^*$ (like W.M. eqs.), this spinorial connection is unusual. It agrees with our conception that the definition of space and time implies non inertia and it raises the question of the status of the more familiar Lorentz spin connection $\delta_A \psi = -A \psi$ $(A=\vec{\sigma}.\vec{A}_\mu dx^\mu)$ which defines, in a spinorial formalism, the traditional transport of tensors. Since $\delta_X$ and $\delta_A$ are incompatible, it is tempting to consider that $\delta_A$ which refers in our opinion to an erroneous view of space time is determined by the condition that it is the one closest to $\delta_X$. One way to define it is to consider the curvature of $\delta=\delta_A-\delta_X$, i.e. the result of the comparison of the transports of $\psi$ along an infinitesimal closed path. An inspection of $\delta^2 \psi = \alpha \psi + \beta \epsilon \psi^*$ ($\alpha$ and $\beta$ being 2-forms) shows that $\beta=0$ (a natural condition outside matter) is equivalent to the absence of torsion, and that $\int<||\delta^2 \psi||^2>$ (exterior product and average on spinors) is Einstein-Hilbert's action with a cosmological constant $\Lambda=a^{-2}$ (if one neglects quadratic terms in the curvature of $A$). \textit{From this point of view, the manifestation of $\Lambda$ in present cosmology is intimately related to a "Newtonian" approach of inertia}. In presence of local matter, $\delta_X$ must be modified, for instance by changing $q$ in $q+Ga^2T^\alpha _\mu q_\alpha dx^\mu$ (on the basis of dimensional arguments and of our knowledge that matter generates non inertia) ; then the cosmological constant looks like a black energy contribution $(Ga^2)^{-1} \eta_{\mu \nu}$ as in Einstein's  equations.

\section{Conclusion : Physics and references, a bird's eye view.}
Since its (unknown) beginning, physics has dealt with the study of phenomena, in particular motions, changes and gravity of matter. Clearly, this study implies the use (or definition) of references ; we recall that the word "phenomenon" like "phase" comes from greek verbs meaning "to appear", "to show". For greeks these references were space and time (the theater), and atoms (the actors), all being absolute ones. They have been associated with the sciences of geometry, astronomy and (in middle ages) alchemy. The first two have generated classical mechanics which confronted with chemistry has led to Q.P (unification of the descriptions of motions and changes of matter). Physicists are presently at work with Q.P and gravity, but no consensus exists, even on the starting point. 
\par Since Galileo and Einstein some time after, neither space nor time is absolute. The references are now inertial frames connected by Poincar\'e transformations. Like Galileo's boats they still are associated with massive bodies (electrons, protons, ..., stars, galaxies or even CMB) ; taking into account gravitation has just given them a local character. Finally, atoms also have lost their absolute character : with the standard model, zero mass particles seem to be the present references for matter. 
\par In our epistemological reflexion on (Galileo-Newton-Einstein-Dirac) past physics, INERTIA has been also the reference, but there it meant "$v=c$" or "light cone" or more fundamentally "spinors" (neutrinos being its best "material" approximation). Physics dealt with NON INERTIA or ACTIVITY (or action). We have successively argued that mass, time, gravitation, space can be thought of in quantum relativistic terms thanks to this concept ; the cosmological constant itself has come out from the definition of space time as an abstract non inertial process. Surely quantum gravity, or the understanding that physics  is not only relativistic and quantum but also  "gravitational", needs additional reflexion and  work. We are conscious that we have put forward ideas but not a new theory. 
\\
\\

Acknowledgments : Some of the ideas suggested in this presentation may be considered as a reformulation of : T. Jacobson and L. S. Schulman's zigzag Feynman's paths to the 1d Dirac eq. ; R. Penrose's introduction of zigzag particles in his "\textit{road to reality}" ; M. Sachs' faith in the role of "\textit{spinors from fermis to light years}". We also thank Prs C. Bracco, F. Debbasch, J.-M. L\'evy-Leblond and B. Raffaelli  for stimulating discussions.

\begin{figure}[hbt]
\includegraphics{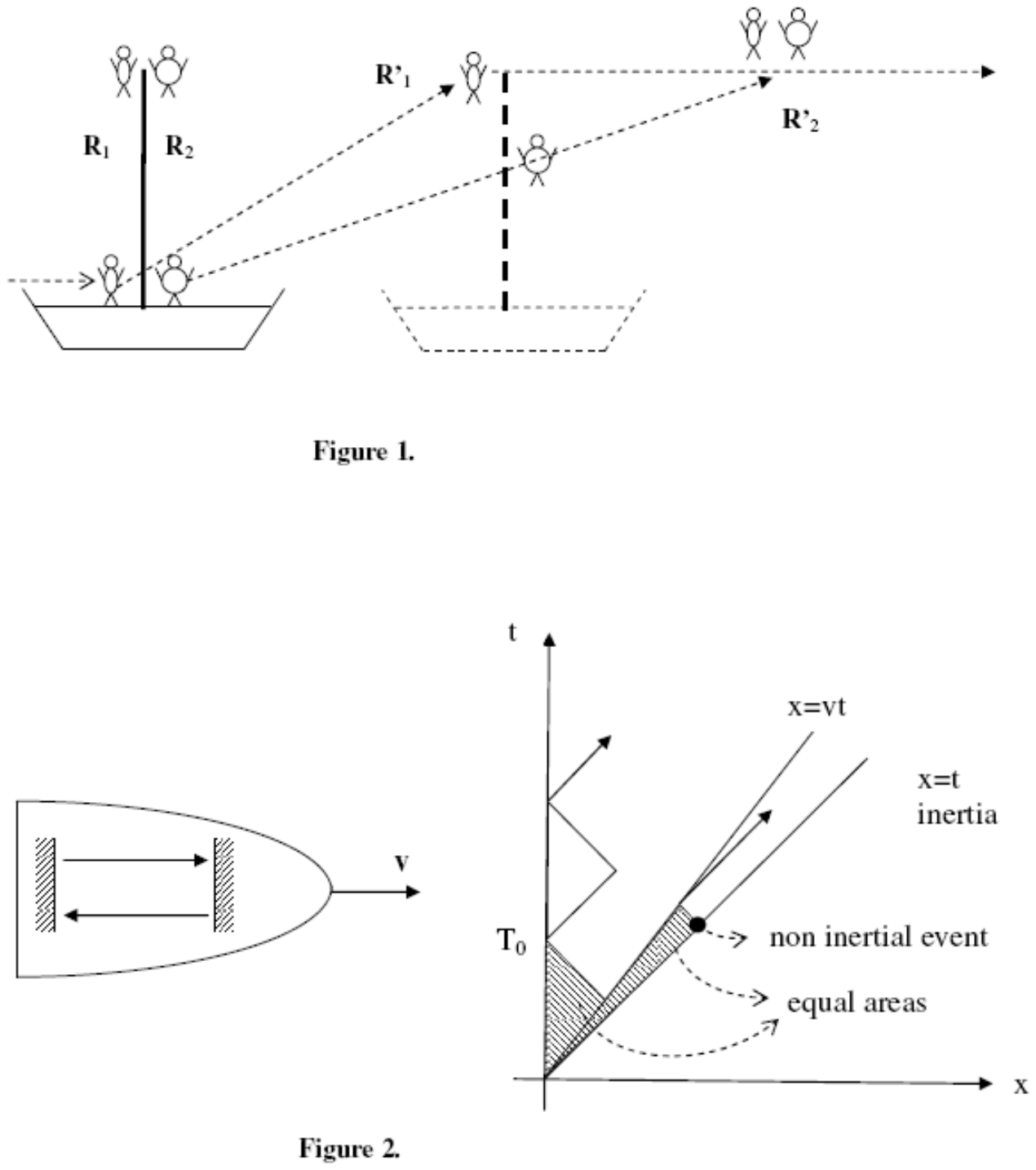}
\end{figure}

\begin{figure}[hbt]
\includegraphics{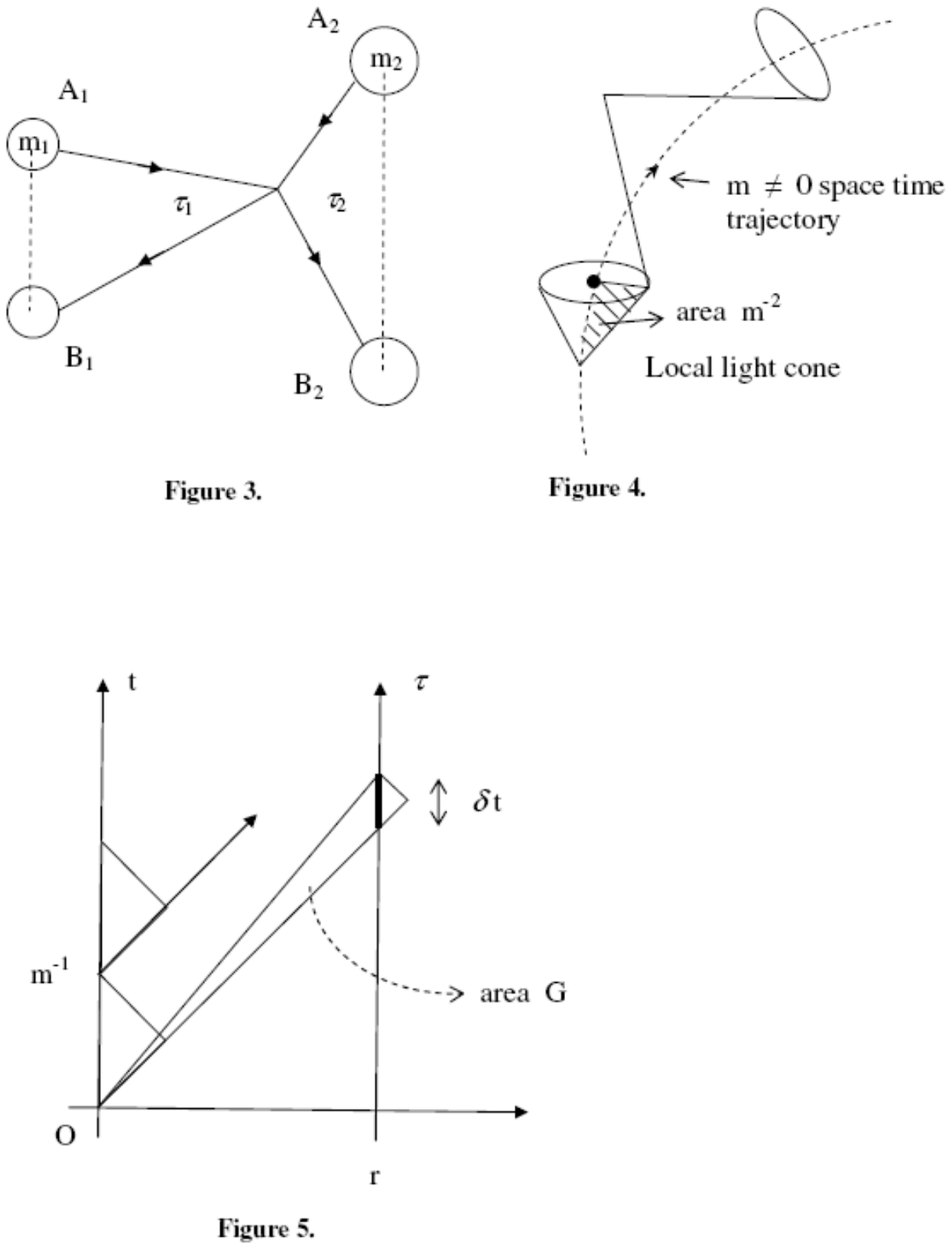}
\end{figure}

\end{document}